\documentclass[reprint,
 amsmath,amssymb,
 aps,prr,
]{revtex4-2}

\usepackage{graphicx}
\usepackage{subcaption}
\usepackage[justification=Justified]{caption}
\usepackage{dcolumn}
\usepackage{bm}
\usepackage{xcolor}
\usepackage{tabularx}
\usepackage{array}
\usepackage{float}
\usepackage{lipsum}
\usepackage{hyperref}
\usepackage{appendix}
\hypersetup{
    pdfborder={0 0 0},
    colorlinks=true,
    linkcolor=black,
    urlcolor=blue,
    citecolor=blue
}
\usepackage{stmaryrd}
\usepackage[none]{hyphenat}

\begin{document}

\preprint{APS/123-QED}

\title{
Machine learning non-Markovian two-level quantum noise spectroscopy
}

\author{Juan M. Scarpetta$^{1}$}
\author{John H. Reina$^{1}$}
\author{Morten Hjorth-Jensen$^2$}

\affiliation{$^1$Departament of Physics and Centre for Bioinformatics and Photonics---CIBioFi, %Edificio E20 No. 1069, 
Universidad del Valle, Cali 760032, Colombia}
%\affiliation{$^2$Departamento de Física, Universidad del Valle, Cali 760032, Colombia}
\affiliation{$^2$Department of Physics and Center for Computing in Science Education, University of Oslo, Oslo N-0316, Norway}

\date{\today}

\begin{abstract}
We develop machine learning models for the automated  characterization of quantum noise spectroscopy for non-Hermitian two-level systems.
We use the Random Forest, Support Vector and  Feed-Forward Neural Network regression algorithms
to perform a highly accurate regression of the two-level system-bath coupling strength. 
High accuracy Ohmicity classification was implemented to provide a complete characterization of the spectral density function.  
We define a time-averaged trace-distance metric to feed the machine learning algorithms which, together with 
numerically exact populations as inputs, produce a highly accurate non-Markovian regression spanning the transition from fast to slow baths and from weak to strong coupling 
regimes of the interaction. The dynamics database of the non-Hermitian systems has been built up within the independent spin-boson 
and pure dephasing 
model.
\end{abstract}

\maketitle

\section{\label{sec1:Introduction} Introduction}

Quantum spectroscopy of dissipative two-level systems  
(TLSs) or quantum bits (qubits) is a key tool for understanding and benchmarking the conditional dynamics that take place in quantum processors~\cite{QCdiscovery,NonlinearSingleSpinSpectrum,QNSnonGaussian,QubitControlNoiseSpectroscopy,SpinQubitNoiseSpectroscopy}, and for the design of quantum noise tolerant networks and computing platforms~\cite{NISQ_Algorithms,QError_Mitigation,QIprotect,QuantumErrorC_AI}.
Quantum technologies, such as quantum computing, quantum sensing and quantum information engineering, rely on the precise control of system dynamics and the exploitation of their quantum properties~\cite{QC_NISQ,QCdiscovery,QCtopological,QCmitigate,QSensing,NHQSensing,QSLsensing,QMetrology,NMQmetrology}. However, quantum systems are inevitably immersed in an environment that induces dissipative effects that  have a non-negligible impact on the dynamics of the system~\cite{QDissipation_Weiss,OpenQuantumS_Petruccione,Review_Markovian,Review_nonMark_Dynamics}. 

To address this challenge, the study of open quantum systems~\cite{QDissipation_Weiss,OpenQuantumS_Petruccione,CaldeiraLeggett_2,JH_QDecoherence,Review_Markovian,Review_nonMark_Dynamics} becomes fundamental for understanding their interaction with their environment, aiming to characterize dissipation phenomena by developing models that describe such complex interactions~\cite{Review_Markovian,JH_QDecoherence,Review_nonMark_Dynamics,CaldeiraLeggett_2,SpinBoson_GenericModel,QInfo_EC}. Several studies have sought to improve our understanding of this process, proposing efficient ways to control its effects~\cite{QIprotect,QError_Mitigation,QCtopological,QCmitigate,QuantumErrorC_AI}. Such manipulations can be achieved, for example, through quantum coherent control techniques~\cite{CoherentControl_TLS_nonMarkovian_ReinaEckel,QIprotect,QuantumCoherentControl} or error correction protocols~\cite{QError_Mitigation,QCtopological,QCmitigate,QuantumErrorC_AI}, which require probing the system to gather information about its environment.

%%%%%%%%%%%%%%%%%%%%%%%%%%%%%%%%%%%%%%%%%%%%%%%%%%%%%%%%%%%%%%%%%%%%%%%%%%%%%%%%%%%%%%%%%%

We use TLSs or qubits as the building blocks of models of open quantum systems that represent the minimal units of information with which we can study the primary effects of a large reservoir in statistical equilibrium on a TLS undergoing relaxation and decoherence. Models such as the spin-boson model~\cite{QDissipation_Weiss,CaldeiraLeggett_2,SpinBoson_GenericModel} are commonly used for this purpose.
%%%%%%%%%%%%
Due to the complexity of the system-environment interactions, such models are often heuristic and oversimplified. For instance, harmonic environments assume that the bath is coupled to the target system within a predefined  frequency  range. Even in these cases, the analysis is rather complicated and requires sophisticated analytical and numerical methods. 

In the Markovian regime, where a large separation of timescales between the system and the environment is required, a key method for describing open quantum systems is the Lindblad master equation.  This simplifies the description by assuming a featureless bath in which all dissipative effects are condensed into jump operators with their respective decay rates~\cite{QDissipation_Weiss,OpenQuantumS_Petruccione}. More robust methods that account for  memory (non-Markovian) effects in the system-environment interactions, where the system and bath timescales  are comparable, include the hierarchical equations of motion (HEOM)~\cite{HEOM_1,HEOM_2}, the quasi-adiabatic path-integral (QUAPI)~\cite{QUAPI_1,Non-Markovian_QUAPI,CoherentControl_TLS_nonMarkovian_ReinaEckel} and the time-evolving matrix product operator (TEMPO)~\cite{TEMPO}. While these methods are numerically exact, they tend to be computationally expensive.

Building on this line of research, we focus on the spin-boson model, which incorporates an interacting bosonic bath of harmonic oscillators and can therefore be characterized by the spectral density (SD) function. 
The SD function can be used to calculate the correlation function~\cite{ExponentialDecomposition_CorrelationFunction} of the environment, determining the time scales of influence, and it encodes all information about the system-bath coupling~\cite{SpinBosonModel_UltraStrong,Non-Markovian_QUAPI}. This SD characterization is based on key parameters such as Ohmicity, cut-off frequency, and coupling strength regimes. However, this function is rarely available for arbitrary quantum systems and is typically inferred from experimental data or imposed by~\textit{ad hoc} assumptions.

%%%%%%%%%%%%%
Despite the fact that the SD function encoding the system-environment interaction is inaccessible in practice unless the system is probed~\cite{ML_ProbingSystem}, one still faces the challenge of \textit{learning} something about an unknown environment~\cite{QInfo_EC} in order to realistically reproduce the system's dynamics. For this reason, in this work we implement machine learning (ML) tools to predict the parameters that fully characterize the properties of a bosonic environment, using the system's time evolution as input. Extracting these parameters allows us to determine the coupling regimes, Ohmicity and non-Markovian nature of the interaction.

Recent works~\cite{Paternostro_1, Paternostro_2} have attempted to investigate dissipative effects in TLSs using neural networks (NNs). They achieved high accuracy in classifying the Ohmicity of an environment through its SD function, by using the time evolution of the system's density matrix under a pure dephasing model. However, crucial parameters influencing the dynamics, such as coupling strength, cutoff frequency and tunneling energy (in the case of spin-boson systems), were not considered during training and classification. In this study, we therefore further explore the parameter space of the environment.
Here we have successfully identified scenarios ranging from fast to slow baths and from weak to strong coupling regimes of interaction, enabling the identification and quantification of non-Markovian dynamics.

Previous developments in quantum technologies explored the usage of machine learning techniques to efficiently solve complex data-driven problems. The intersection of ML and quantum physics has been extensively exploited in various fields~\cite{NISC_Control&RL, ML_Control_QuantumSystems, Status_QML,Marquardt2023}. These methods include reinforcement learning for optimal coherent control~\cite{OCC_ReinforcementLearning}, quantum thermodynamics~\cite{RL_Thermodynamics}, quantum circuits for modeling non-Markovian dynamics~\cite{QuantumComputing_nonMarkovian} and NNs for simulating the quantum dynamics of open quantum systems~\cite{NN_Simulating_Quantum_Dynamics}.  The work of Krenn {\em et al.}~\cite{Marquardt2023} has several interesting perspectives on machine learning and quantum technologies. 

While NNs are the most widely used method due to their adaptable architecture, which can easily be customized for any regression or classification task, this work investigates alternative ensemble methods for regression, such as the Random Forest Regressor (RFR) and the Support Vector Regressor (SVR). These methods also achieve high precision when predicting parameters from complex inputs. Additionally, the metrics used to define various prediction measures are analyzed, including accuracy, mean squared error (MSE), $R^2$-score, recall and $F1-$score.  These are widely used in such tasks~\cite{ML_metrics,Hastie2009,Murphy2012}.
 
Several metrics have been proposed to determine the non-Markovianity of an arbitrary quantum evolution process, $\Phi_t$. Each  metric depends on the physical context (e.g. quantum computing, quantum optics, and few-level systems)~\cite{Review_nonMark_Dynamics}. Quantifying the non-Markovianity of a system requires precise metrics, such as BLP~\cite{BLP_MeasureDegreeNon-MarkovianBehavior,BLP_measure_Original} and  RHP~\cite{nonMark_RHP}, among others. These  quantify the degree of non-Markovianity, but depend on the choice of initial states~\cite{BathInitial_Conditions} or are difficult to compute. Furthermore, these metrics do not always agree, as they capture different aspects of non-Markovianity; they are based on different physical properties, such as information, entanglement, and divisibility, all of which allow memory in quantum systems to be quantified.

Here, we report the successful prediction of the direct, automated characterization of the degree of non-Markovianity of $\Phi_t$ using ML methods with different labeled targets, thus avoiding the need for lengthy and complex computations of these metrics.

The paper is structured as follows: Section~\ref{sec2:TheorMethd} presents the theoretical framework, including the spin-boson and  pure dephasing models.
Section~\ref{sec:Methods} describes the methods used, particularly the dataset generated to train and evaluate the ML algorithms. It also briefly outlines the  Random Forest, Support Vector Regressor and Feed-Forward Neural Networks (FFNN) algorithms, the technical details of which are provided in Appendix (Section~\ref{sec:Methods_ML}). Sections~\ref{sec:Results} and \ref{sec:Discussion} present the main results and discussion. Finally, Section~\ref{sec:conclusion} presents  the conclusions. Some of the code that we developed has been open-sourced.

\section{\label{sec2:TheorMethd} Theoretical Framework}

We analyze non-Hermitian open quantum systems describing  dissipative processes in which a TLS interacts with a reservoir. 
We consider the paradigmatic
spin-boson model, which captures the interactions between the system and the reservoir~\cite{CaldeiraLeggett_2}, in order to construct ML models capable of automating  non-Markovian quantum spectroscopy of TLSs.  

 Two main processes are identified for the interacting dissipative TLS: decoherence, where the phase of the TLS state is randomized by its entanglement with the bath without energy exchange~\cite{JH_QDecoherence}, and damping or relaxation, where the TLS exchanges energy with its surrounding environment, eventually reaching thermal equilibrium~\cite{CaldeiraLeggett_2,Review_Markovian,Review_nonMark_Dynamics}.  Mathematically, for quantum systems exhibiting the latter process, the TLS operator does not commute with the total Hamiltonian, making it a non-Hermitian system~\cite{Review_Markovian,Review_nonMark_Dynamics,JH_QDecoherence,CaldeiraLeggett_2,SpinBoson_GenericModel}. On the other hand, the former 
process alone defines another type of non-Hermitian system in which the TLS operator commutes with the total Hamiltonian: there is a loss of coherence in the TLS state, but the total energy of the system remains constant over time. This scenario is often referred to as the pure dephasing model.
%%%%%%%%%%%%%%%%%%%%%%
\begin{figure}[h!]
    \centering
\includegraphics[width=0.8\linewidth
]{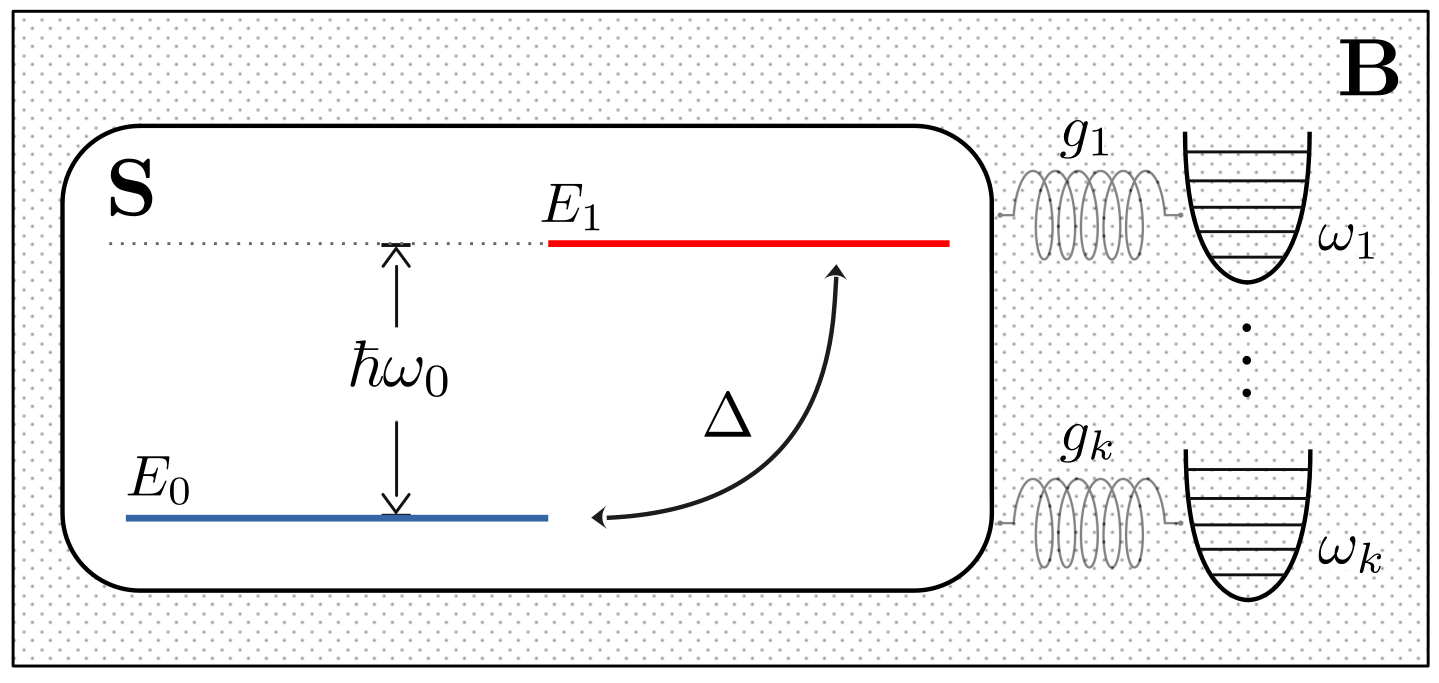}
    \caption{Sketch of a generic TLS  $\mathbf{S}$ of transition frequency $\omega_0$ and tunneling energy $\Delta$, coupled to a bosonic bath $\mathbf{B}$ of infinitely many oscillators. The coupling constants $g_k$ define the S-B interaction strength of each mode $\omega_k$.}
\label{fig:Sketch_SBModel}
\end{figure}
%%%%%%%%%%%%%%%%%

We use the spin-boson model to analyze the physics of described non-Hermitian open quantum systems. 
Consider a quantum system $\mathbf{S}$ with two energy levels $E_0$ and $E_1$ and a transition frequency $\omega_0$. 
The system $\mathbf{S}$  interacts with a bath $\mathbf{B}$, leading to dephasing effects and relaxation between the energy levels~\cite{Review_Markovian,Review_nonMark_Dynamics,JH_QDecoherence,CoherentControl_TLS_nonMarkovian_ReinaEckel,CaldeiraLeggett_2,SpinBoson_GenericModel}. 
The bath is modeled as a large set of quantum harmonic oscillators, where the $k$-th oscillator couples to the system at frequency $\omega_k$ through a coupling constant $g_k$, and can be described by the so-called \textit{bosonic bath} Hamiltonian
$ H_B = \sum_k \hbar\omega_k(\hat{a}^{\dagger}_k\hat{a}_k + \frac{1}{2})$,
where $\hat{a}^{\dagger}_k$ ($\hat{a}_k$) are the creation (annihilation) operators of mode $k$, which satisfy the usual bosonic commutation relations
~\cite{QDissipation_Weiss,OpenQuantumS_Petruccione, SpinBoson_BosonicBath}.
Figure~\ref{fig:Sketch_SBModel} illustrates the relevant energies and system-bath (S-B) couplings.

The total Hamiltonian describing the system is given by $H =H_S+H_B+H_{SB}$, where 
$H_{S}$ denotes the TLS Hamiltonian. The interaction Hamiltonian, $H_{SB}$, is described as a linear coupling between the system operator, $\hat{Q}$, and the bath operator, $\hat{B}$, as follows:
%%%%%%%%%%%%%%%%
\begin{eqnarray} 
H_{SB} = \hbar\,\hat{Q}\otimes\hat{B}.
\label{Eq:H_sb} 
\end{eqnarray}
%%%%%%%%%%%% 
In turn, $B$ 
is expressed as the sum of the creation and annihilation operators of each mode: $\hat{B}= \sum_k (g_k \hat{a}^{\dagger}_k + g_k^*  \hat{a}_k)$, and the TLS-bath coupling can be written as
%%%%%%%%%%%%%%%%%%%%
\begin{equation}
    H_{SB}  = \frac{1}{2}\hbar\hat{\sigma}_z\otimes  \sum_k \left(g_k \hat{a}^{\dagger}_k + g_k^*  \hat{a}_k   \right).
    \label{Eq:H_sb1}
\end{equation}
%%%%%%%%%%%%%

The TLS coupling with bath modes $k$ is characterized by an SD function $J(\omega) = \sum_k |g_k|^2 \delta(\omega-\omega_k)$.
Typically, $k$ denotes  a much larger number of  degrees of freedom than the TLS, meaning that $J$ becomes a continuous function of the frequency $\omega$.
Different types of spectral densities have been studied~\cite{CaldeiraLeggett_2,SpinBoson_GenericModel,JH_QDecoherence,Non-Markovian_QUAPI,SD_markovians,SpinBoson_Decoherence,UnderDamped_SpinBoson_HEOM,MolecularSDreconstruction}, each with different properties and effective frequency regimes. 
Several analytical and numerical techniques can be employed to calculate the dissipative time evolution of the TLS~\cite{QDissipation_Weiss,OpenQuantumS_Petruccione,Review_Markovian,CaldeiraLeggett_2,Review_nonMark_Dynamics,JH_QDecoherence}.

We consider a general SD function of the form
%%%%%%%%%%%%%%%%
\begin{equation}
    J_s(\omega) = \eta \omega_c^{1-s} \omega^s \mathrm{e}^{-\omega/\omega_c} ,
    \label{Eq:Js(w)}
\end{equation}
%%%%%%%%%%%%%%%%
with an exponential cut-off at frequency $\omega_c$, where $\eta$ is the system-bath coupling strength and $s$ is the ohmic property of the bath, usually expressed as 
\textit{Class 0:} $0 < s < 1$, sub-Ohmic;
\textit{Class 1:} $s=1$, Ohmic and
\textit{Class 2:} $s > 1$, super-Ohmic. 

%%%%%%%%%
Another type of spectral density considered here is the Lorentz-Drude  function
%%%%%%%%%%%%%%%%
\begin{equation}
    J_L(\omega) = \frac{2\gamma \omega_c \omega}{\omega^2+\omega_c^2},
     \label{Eq:DL}
\end{equation}
%%%%%%%%%%%%%%%%%
where $\gamma$ is the coupling strength energy  between the system and the bosonic environment with bandwidth $\omega_c$. This SD function is used during the dataset generation of the non-Markovian dynamics in the spin-boson model (cf. Sec.~\ref{sec:Dataset}), as it provides a convenient  implementation of the Matsubara exponential decomposition in the correlation function. 
Furthermore, by appropriately fine-tuning the parameters $s$ and 
$\eta$ in Eq.~\eqref{Eq:Js(w)},
%$\gamma$, 
it is possible to recover Eq.~{\eqref{Eq:DL}~\cite{UnderDamped_SpinBoson_HEOM}.
Figure~\ref{fig:SD_functions} shows a comparison between  $J_s(\omega)$ and $J_L(\omega)$. 

The quantum dissipative dynamics associated to such SD functions is plotted in Fig.~\ref{fig:Dataset}.  
The construction of machine learning models for the characterization of TLSs quantum noise spectroscopy using these Ohmic-type and Lorentz-Drude spectral densities will be considered below.
%%%%%%%%%%%%%%%%%%%
\begin{figure}
    \centering
\includegraphics[width=0.8\linewidth]{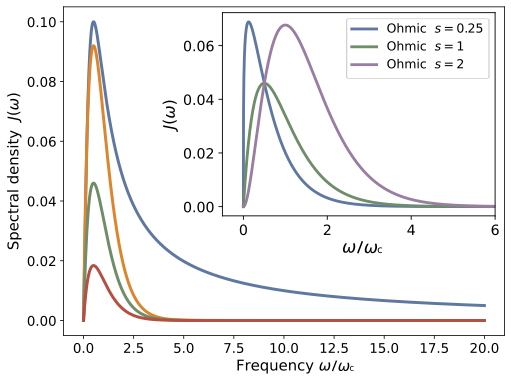}
\caption{Comparison of different spectral density functions. The Lorentz-Drude SDF with $\gamma = 0.1$ and $\omega_c = 0.5$ is shown in blue. The orange, green and red curves represent Ohmic SDFs $J_{1}(\omega)$ with $\eta=0.5$, $\eta=0.25$, and $\eta=0.1$, respectively, and cut-off $\omega_c=0.5$. The inset shows the SDFs from Eq.~\eqref{Eq:Js(w)} for fixed $\omega_c=0.5$, $\eta=0.25$ and different $s$ values. }
\label{fig:SD_functions}
\end{figure}
%%%%%%%%%%%%%%%%%%%

%%%%%%%%%%%%%%%%%%%%
\begin{figure*}[ht]
\centering
\includegraphics[width=0.9\linewidth]{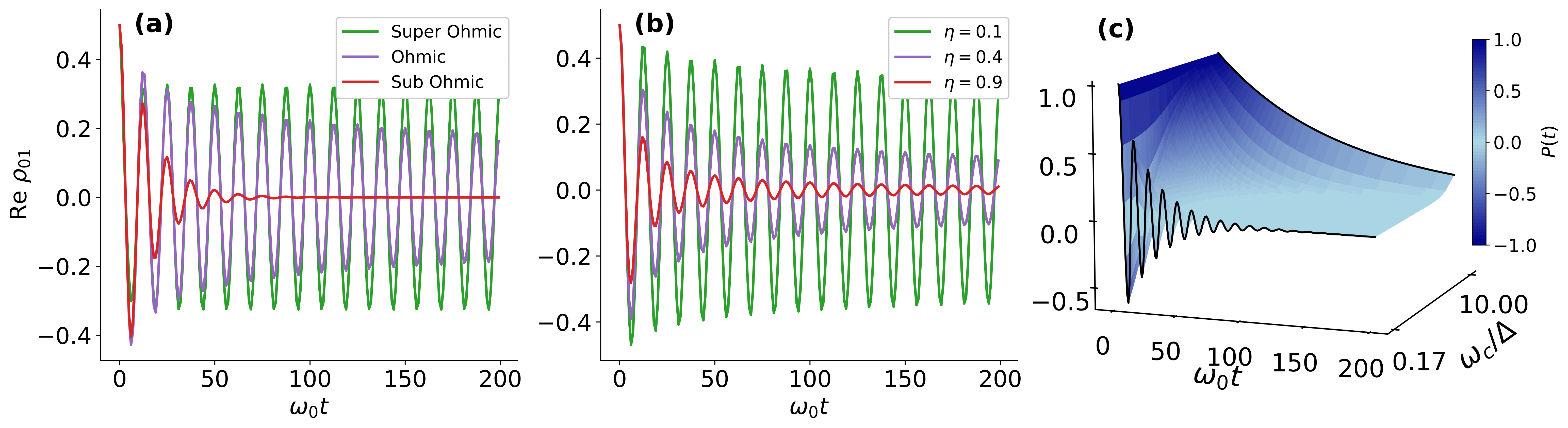}
    \caption{Case $\Delta= 0$: Dynamics of the real part of the coherence $\text{Re }\rho_{01}$ as a function of Ohmicity $s$ and coupling strength $\eta$; (a) sub-Ohmic ($s=0.1$), Ohmic ($s=1$) and super-Ohmic ($s=3.5$) spectral densities;  $\omega_c=0.5$, $\eta=0.25$ and $\omega_0=1$;    
    (b) Ohmic dynamics %($s=1$) 
    for $\eta=0.1, 0.4, 0.9$, $\omega_c=0.5$ and $\omega_0=1$. In both cases, Eq.~\eqref{Eq:rhoS_analytical} was used with initial condition $\hat{\rho}_S(0)= |+\rangle \langle+|$ at $T=0$. (c) Case $\Delta\neq 0$: TLS population difference $P(t)$ as a function of time and the ratio of the bath bandwidth to the tunnelling energy, $\omega_c/\Delta$ ($\hbar=1$). The numerically exact calculation allows the full transition from fast (Markovian) to slow (non-Markovian) baths; $\gamma = 0.25$, $\omega_0=1$, $k_\text{B}T = 0.5 \Delta$ and TLS initial state $\hat{\rho}_S(0)= |1\rangle \langle1|$.
    }
\label{fig:Dataset}
\end{figure*}
%%%%%%%%%%%%%%%%
\subsection{Pure dephasing model \label{sec:PureDephasing}
}
As there is no energy exchange between the TLS and the bath, the system's populations remain unchanged. Coherence decay can therefore be calculated exactly, since the TLS operator $\hat{Q}\equiv\hat{\sigma}_z$ %(see Eq.~\eqref{Eq:H_sb}) 
commutes with the total Hamiltonian $H$. The coupling Hamiltonian, $H_{SB}$, is given by Eq.~\eqref{Eq:H_sb1} and  
the TLS Hamiltonian~\cite{OpenQuantumS_Petruccione,JH_QDecoherence}  
%%%%%%%%%%%%%%%%%%%%
\begin{eqnarray}
    H_S = \frac{1}{2}\hbar \omega_0 \hat{\sigma}_z. 
    \label{Eq:H_s}
\end{eqnarray}
%%%%%%%%%%%%%
The analytical solution for the TLS reduced density matrix is given by
%%%%%%%%%%%%%%
\begin{equation}
    \hat{\rho}_S(t) = \left(
    \begin{array}{cc}
        \rho_{00} & \rho_{01}e^{-\Gamma(t)+i\omega_0t}\\
        \rho_{01}^*e^{-\Gamma(t)-i\omega_0t} & 1-\rho_{00}  \label{Eq:rhoS_analytical}
    \end{array}
    \right),
\end{equation}
%%%%%%%%%%%%%%
where $\Gamma(t)$ is the characteristic decoherence function~\cite{OpenQuantumS_Petruccione,JH_QDecoherence}
%%%%%%%%%%%%%%%%%%
\begin{equation}
    \Gamma(t) = 4\int_0^\infty \text{d}\omega\; J_s(\omega)\coth\left(\frac{\hbar \omega}{2k_B T}\right) \frac{1-\cos(\omega t)}{\omega^2}  .
    \label{Eq:DecohFunction}
\end{equation}
%%%%%%%%%%%%%%%%%%%%%%
The values of $\rho_{00}$ correspond to the initial population of the ground state $| 0 \rangle$, while $\rho_{01}$ to the initial coherence, which is determined by the initial condition   $\hat{\rho}_S(t=0)$.
%%%%%%%%%
Since for this model the populations of the system remain constant, the relevant information lies in the coherences, which exhibit damped oscillations over time depending on the bath parameters $s$, $\eta$ and $\omega_c$. Examples of damped coherence oscillations are shown in Figures~\ref{fig:Dataset}(a) and ~\ref{fig:Dataset}(b), where the SD function Eq.~\eqref{Eq:Js(w)} has been used. This model is particularly useful for studying several physical systems, since decoherence usually occurs much faster than relaxation processes. It also provides an analytical solution for the decoherence function. For instance, it has recently been employed to identify basic chemical principles behind electronic quantum
decoherence of molecules in the condensed phase~\cite{MolecularSDreconstruction,MolecularRoomTQC,MolecularUltrafastQC_Solvent}.

\subsection{\label{sec:SB-model}Spin-boson model
}

The interacting quantum system $\mathbf{S}$ (see Fig. \ref{fig:Sketch_SBModel}) is   described more generally by the Hamiltonian~\cite{CaldeiraLeggett_2}
%%%%%%%%%%%%%%%%%%%%
\begin{equation}
    H_S = \frac{1}{2}\hbar \omega_0 \hat{\sigma}_z + \Delta \hat{\sigma}_x ,
    \label{Eq:H_s1}
\end{equation}
%%%%%%%%%%%%%%%%%%%%
where $\Delta$ denotes the tunnelling energy and 
$\hat{\sigma}_{x,z}$ 
are the Pauli matrices. The interaction  Hamiltonian is given by  Eq.~\eqref{Eq:H_sb1}; therefore $H_S$ (Eq.~\eqref{Eq:H_s1}) does not commute with the  $H_{SB}$ operator, and a general analytical expression for the $\hat{\rho}_S$ components is no longer available. In this case, oscillations in the coherences and changes in the populations also occur, and numerically exact algorithms such as HEOM~\cite{HEOM_1,HEOM_2} or QUAPI~\cite{QUAPI_1,CoherentControl_TLS_nonMarkovian_ReinaEckel,Non-Markovian_QUAPI} can be employed to calculate the dynamics. 

Throughout the dynamics, the bosonic bath remains in thermal equilibrium at temperature $T$ and can be  described by a Gibbs thermal state 
$\hat{\rho}_B =\frac{1}{Z_B} \mathrm{e}^{-\beta H_B}$, 
with associated bath partition function $Z_B = \text{Tr} (\mathrm{e}^{-\beta H_B})$. For the initial condition at $t=0$, we consider the separable state $\hat{\rho}(0) = \hat{\rho}_S(0) \otimes \hat{\rho}_B$. In the interaction picture, the $\hat{B}$ operator 
%in Eq.~\eqref{Eq:B(t)} 
 transforms as $\mathrm{e}^{iH_Bt}\hat{B} \mathrm{e}^{-iH_Bt}$ to obtain
$\hat{B}(t) = \sum_k (g_k \hat{a}^{\dagger}_k \mathrm{e}^{i\omega_kt}+ g_k^*  \hat{a}_k \mathrm{e}^{-i\omega_kt})$,
with bath autocorrelation function $C(t)=\text{Tr}\lbrace \hat{B}(0) \hat{B}(t) \hat{\rho}_B \rbrace$. We obtain
%%%%%%%%%%%%%%%%%
\begin{equation}
    C_\beta (t) = \int_0^\infty J(\omega) \left[ \cos(\omega t)\coth\frac{\beta \hbar\omega}{2} -i \sin(\omega t) \right] \text{d}\omega, %\nonumber 
    \label{Eq:CorrFunct}
\end{equation}
%%%%%%%%%%%%%%%
where $\beta=1/k_\text{B} T$ is the inverse temperature, $k_\text{B}$ is the Boltzmann constant, and $k_\text{B}T$ is the thermal energy.

 $C_\beta (t)$ enables all bath correlations relevant to the system to be determined once its spectral density function, $J(\omega)$, has been established.  In the numerically exact results reported in Fig.~\ref{fig:Dataset}(c), we set $J(\omega)\equiv J_L(\omega)$. 
The bandwidth $\omega_c$ sets the timescale for the bath correlation dynamics $\tau_c\equiv 1/\omega_c$. The reorganization energy $E_r = 2\gamma$, as obtained from Eq.~\eqref{Eq:DL}, where 
$\gamma$ is the S-B coupling strength energy.
The full structure of $C_\beta(t)$ (including both the real and imaginary parts) has been incorporated into 
the HEOM method employed here to compute the TLS dynamics,  thus enabling a complete non-Markovian description of the system dynamics. Figure~\ref{fig:Dataset}(c) plots the TLS population difference,
$P(t)\equiv \rho_{11}(t)-\rho_{00}(t)$, which illustrates the transition from a fast to a slow bath within the strong coupling regime of the system-bath interaction.

\section{Methods\label{sec:Methods}}

\subsection{Dataset generation\label{sec:Dataset}}
In order to identify and classify the effect of the bath on the dynamics of the system, the time evolution of the density matrix components of the system was calculated for different parameter values. We first consider 
the pure dephasing scenario ($\Delta=0$), with reduced density matrix components given by Eq.~\eqref{Eq:rhoS_analytical}, and set $T=0$. The real part of the coherence $\rho_{01}$ is then considered as the dephasing information. The Ohmicity and coupling strength regime are completely determined by the parameters $s$ and $\eta$, respectively. In all cases, an evolution of 200 time steps was taken.  Therefore, a first scenario is created where both regimes are relatively easy to distinguish, as the distribution of these parameters is well separated. This creates a first classification criterion that is easy to distinguish for machine learning methods.

For the classification of the parameter $s$, in the scenario with separated data, 10 000 different values were generated in the intervals $s\in(0.1,0.25)\cup\lbrace1\rbrace\cup(2, 4)$, while for the scenario with non-separated data, 10 000 values were also taken, uniformly distributed in the interval $s\in(0.1, 4)$. On the other hand, for the regression of the $\eta$ parameter, a logarithmic distribution was used in the intervals $\eta \in (10^{-2}, 10^{-1})\cup (0.7, 1)$ for the separated data. In contrast, for the continuous data, a distribution of 10 000 values was generated in the interval $\eta \in (10^{-3}, 1)$. Figures~\ref{fig:Dataset}(a) and~\ref{fig:Dataset}(b) show some curves for different values of $s$ and $\eta$, calculated from  Eq.~\eqref{Eq:rhoS_analytical}, under the initial condition $\hat{\rho}_S(0)= |+\rangle \langle+|$,  $|+\rangle=\frac{1}{\sqrt{2}} (|0\rangle +  |1\rangle $).

For the case $\Delta \neq 0$, a more complex dynamics arises, in which both populations and coherences change over time. 
The characteristic timescale due to the bath response  $\tau_c$, the TLS energy $\Delta$, and the strength coupling  $\gamma$ introduce a  set of parameters that define dynamical Markovian and non-Markovian regimes~\cite{CoherentControl_TLS_nonMarkovian_ReinaEckel, SpinBoson_OscillationsNonMark}. These can lead to population difference showing damped oscillations over time or incoherent exponential decay. The ratio $\omega_c/\Delta$, together with the TLS-bath coupling strength $\gamma$, defines the transition from a fast to a slow bath and from weak to strong coupling regimes of the S-B 
interaction. The population difference
$P(t)$
is plotted as a function of time and of the ratio $\omega_c/\Delta$  in Fig.~\ref{fig:Dataset}(c) for $\gamma = 0.25$. This shows the full cross-over from a Markovian to a non-Markovian bath, generated using 100 values for such a  ratio. The calculations  were performed at high temperature, $k_\text{B}T = 0.5 \Delta$, using the numerically exact HEOM technique.  For the
regression of the $\omega_c/\Delta$ parameter, a
distribution of 10 000 values was generated in the interval  $\frac{\omega_c}{\Delta}\in [0.166, 10.0]$.

\subsection{Non-Markovianity identification and quantification\label{subsec:Methods_ML}}

For the purposes of this discussion, we introduce the ratio of the bosonic bath bandwidth to the tunnelling energy,  $\alpha\equiv\omega_c/\Delta$, a parameter that captures two fundamental properties of the system under consideration.
%%%%%%%%%%%
In the asymptotic limit $\Delta \to 0$, the system can be treated within the pure dephasing model (Eq.~\eqref{Eq:rhoS_analytical}), yielding flat population inversion curves at the base of the surface plot (see Fig.~\ref{fig:Dataset}(c)). However, for tunnelling energies in the scaling limit $\Delta \ll \omega_c$, the population difference exhibits exponential decay over time for  finite $\alpha$, as is clearly shown in Fig.~\ref{fig:Dataset}(c). This monotonic decay behavior indicates an \textit{irreversible} energy flow from the system to the environment, characteristic of a Markovian process~\cite{CaldeiraLeggett_2,Review_nonMark_Dynamics}. This reference dynamics, denoted $\hat{\rho}_S^{\text{ref}}(t)$,  is used to determine the (non-)Markovianity of other processes with different $\alpha$ values. To directly compare two dynamics with different labels, it is necessary to quantify the distinguishability between their states, described by their respective density matrices. 
%%%%
%%%%%%%%%%%%%%%%%%%%
\begin{figure}[t]
    \centering
    \includegraphics[width=0.75\linewidth
    ]{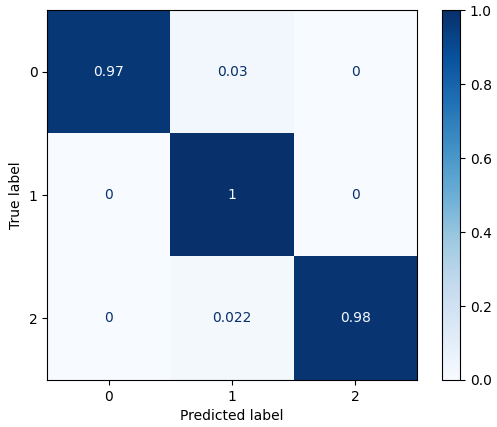}
    \caption{Normalized confusion matrix showing the results of Ohmicity classification using a FFNN on the test set with non-separated data.}
\label{fig:ConfusionMatrix}
\end{figure}
%%%%%%%%%%%%%%%%%%%%%%%%

%%%%%%%%%%%%%%%%%%%%%%
\begin{figure*}[t]
    \centering
\includegraphics[width=0.75\textwidth]{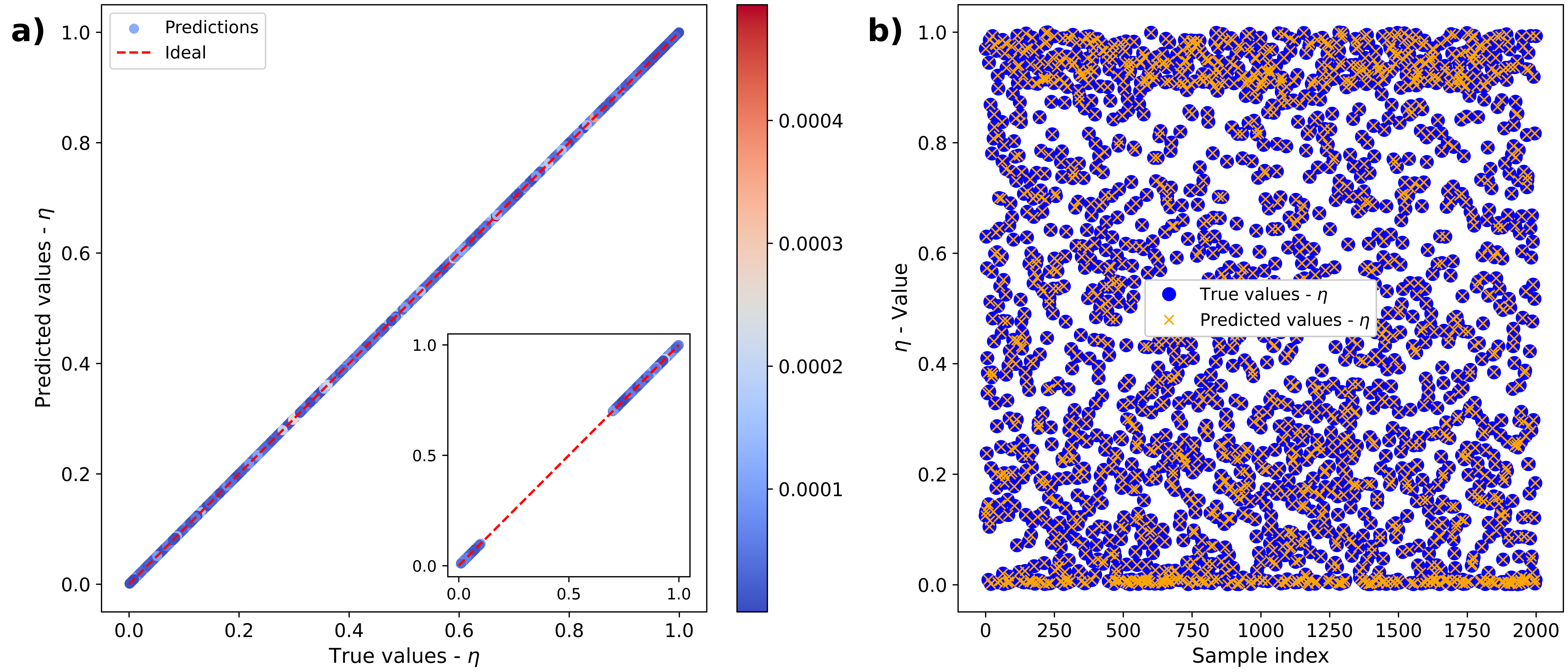}
    \caption{(a) Results of the $\eta$ regression using RFR showing the comparison between predicted and true values. The inset shows the regression for separated data and the colorbar shows the absolute error. (b) The contrast between actual values and model predictions is shown over the parameter space of the test set.}
\label{fig:RFR_results_eta}
\end{figure*}
%%%%%%%%%%%%%%%%%%%
 
The distinguishability between two states $\hat{\rho}_1(t)$ and $\hat{\rho}_2(t)$ at time $t$ can be defined using  the trace distance metric~\cite{NielseChuang_QI_QC}
%%%%%%%%%%%%%%
\begin{equation}
    D_{12}(t) = \frac{1}{2} \text{Tr} \left|\hat{\rho}_2(t)-\hat{\rho}_1(t)\right| ,
\label{eq:TraceDistance}
\end{equation}
%%%%%%%%%%%%%%
where the  norm operator $|\hat{A}|:=\sqrt{\hat{A}^\dagger \hat{A}}$. This metric satisfies $0\leq D \leq 1$ and exhibits properties relevant to the analysis of non-Markovianity,  including contractivity under completely positive maps~\cite{CPM_nonMark, Non-MarkovianQMapsQubitNoise}.
To measure the global similarity between two dynamics up to time $T$, we define a parameter $\sigma$ as the time-averaged $D(t)$:
%%%%%%%%%%%%%%%%
\begin{equation}
    \sigma = \frac{1}{T} \int_0^T D_{12}(t) \;\text{d}t. \label{eq:sigma_definition}
\end{equation}
%%%%%%%%%%%%%%%
By setting the reference dynamics to~$\hat{\rho}_{\text{ref}}=\hat{\rho}_1$ (corresponding to the Markovian regime with $\alpha=10$), we reparametrize the remaining dynamics in terms of $\alpha$ for a fixed $\gamma=0.25$ and $\omega_c=0.5$ as
%%%%%%%%%%%%%%%%
\begin{equation}
\sigma_\alpha = \frac{1}{T} \int_0^T \frac{1}{2}\; \text{Tr} \left|\hat{\rho}_\alpha(t)-\hat{\rho}_{\text{ref}}(t)\right| \;\text{d}t. \label{eq:Sigma_alpha}
\end{equation}
%%%%%%%%%%%%%%%
These $\sigma_\alpha$ values define new labels for the regression problem, encoding the departure from the Markovian scenario by sweeping $\Delta$ through the regimes $\gamma\ll \Delta$ and $\gamma\sim\Delta$.

\subsection{Machine Learning models\label{subsec:Methods_ML}}

Using all this information as input, ML models are built to classify and predict the Ohmicity, coupling strength regime and non-Markovianity of the system dynamics.
We have used the Feed Forward Neural Network algorithm for both classification and regression problems. In addition, we have implemented the Random Forest Regressor and Support Vector Regressor models. The implementation of these methods for the automated characterization  of quantum noise spectroscopy developed here  is described in detail in the Appendix (Sec.~\ref{sec:Methods_ML}).

%%%%%%%%%%%%%%%%%%%%%%%%%%
\section{Results\label{sec:Results}}
%%%%%%%%%%%%%%%%%%%%%%%%%%%

For the classification of the Ohmicity in the dynamics, after the neural network training process, it was expected that the FFNN would perform well in the case of separated data, as the curves are easy to distinguish. As shown in Fig.~\ref{fig:Dataset}(a), the sub-Ohmic dynamics show a faster and more pronounced decay, while the super-Ohmic dynamics show a slower oscillatory behavior. In the Ohmic case, an intermediate region between these two behaviors is observed. After about  200 epochs, 100\% accuracy is achieved on the test data, confirming the excellent performance of the model in this pilot classification task.

In a second step, the training process was repeated for the case of continuous and uniformly distributed values of $s$, ensuring a smooth transition between the Ohmicity categories. After 3000 epochs, approximately 98\% accuracy was achieved on the test data. The classification results are shown in Figure~\ref{fig:ConfusionMatrix} and Table~\ref{tab:OhmicityScores_nonSep}.
%%%%%%%%%%%%%%%%%%%%%
%%%%%%%%%%%%%%
\begin{table}[h]
\centering
\renewcommand{\arraystretch}{1.3}
\begin{tabular}{>{\centering\arraybackslash}p{0.2\linewidth}>{\centering\arraybackslash}p{0.25\linewidth}>{\centering\arraybackslash}p{0.25\linewidth}>{\centering\arraybackslash}p{0.25\linewidth}} % Adjust column widths manually
\hline
\multicolumn{1}{c}{\textbf{Class}}   & \textbf{Precision} & \textbf{Recall}   & \textbf{F1-Score }\\ \hline
0                           & 1.000000  & 0.970266 & 0.984909 \\
1                           & 0.951359  & 1.000000 & 0.975073 \\
2                           & 1.000000  & 0.978448 & 0.989107 \\ \hline
\multicolumn{1}{c}{\textbf{Average}} & \textbf{0.983827}  & \textbf{0.983000} & \textbf{0.983099} \\ \hline
\end{tabular}
\caption{Metrics for the $s$-classification of the FFNN for non separated data.}
\label{tab:OhmicityScores_nonSep}
\end{table}

Moving on to the regression of the parameter $\eta$, the results of the RFR are shown in Figure~\ref{fig:RFR_results_eta}. We can see how the model predicts the values of $\eta$ based on the input features. As expected, the results for the distinction of the strong and weak coupling regimes gave the better $R^2$-score and the lowest Mean Squared Error (MSE) of $6.227\times10^{-10}$ given by the very well separated data. For the regression with uniformly distributed values, we obtained an $R^2-$score of 0.9996 and an MSE of $6.0258\times 10^{-9}$, which is an order of magnitude higher than the separated case. These results highlight the robustness of the model even in scenarios with overlapping parameter distributions.

In order to capture the sensitiveness of the overall dynamics with respect to the coupling strength parameter, we use different types of kernels in the SVR to map the temporal features onto different feature spaces in order to reduce the complexity of the dataset and simplify the regression of the data. Table~\ref{tab:etaScores_SVR} shows the metrics of the SVR regression for different kernels. 
%%%%%%%%%%%%%%%%%
\begin{table}[h!]
\centering
\renewcommand{\arraystretch}{1.3}
\begin{tabular}{>{\centering\arraybackslash}p{0.2\linewidth}>{\centering\arraybackslash}p{0.25\linewidth}>{\centering\arraybackslash}p{0.25\linewidth}>{\centering\arraybackslash}p{0.25\linewidth}}
\hline
\multicolumn{1}{c}{\textbf{Kernel}}   & \textbf{MAE} & \textbf{MSE}   & \textbf{$R^2-$score}\\ \hline
Linear                         & 0.0642568  & 0.00498204 & 0.9576 \\
RBF                            & 0.0579192  & 0.00419503 & 0.9643 \\
Poly                           & 0.0646946  & 0.00502820 & 0.9572 \\ \hline
\end{tabular}
\caption{Metrics of the $\eta$-regression using SVR.}
\label{tab:etaScores_SVR}
\end{table}
%%%%%%%%%%%%%%%%%%

For the FFNN regression model, it was trained using the gradient descent with Adaptive Momentum (ADAM) method  described in Eq.~\eqref{eq:GD_ADAM} to obtain the model for predicting the values of $\eta$. In Figure~\ref{fig:TrainingProcess_eta}, it can be observed that after a total of 500 epochs, an effective minimization of the cost function given by Eq.~\eqref{eq:MSE_loss} is achieved.
\noindent
After the training phase, the model was tested on the test set resulting in a final MSE of $3.3244\times10^{-7}$ for separated $\eta$ data and $1.2355\times10^{-7}$ for non separated data.
%%%%%%%%%%%%%%
\begin{figure}[t]
    \centering
    \includegraphics[width=0.8\linewidth
    ]{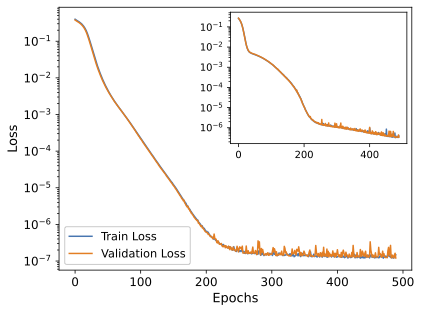}
    \caption{Training process of the FFNN using the ADAM optimizer with a learning rate $\lambda=10^{-5}$ and a batch size $\text{Bs}=64$ to perform $\eta-$regression tasks with non-separated data. Inset: Loss evolution for the separated data case.}
\label{fig:TrainingProcess_eta}
\end{figure}

For the non-Markovian regime of the dynamics, the dataset generated with the population differences $P(t)$ was used.  These components were obtained using the HEOM method, which captures the dynamics for cases of strong coupling and bidirectional flow of information and energy from the bath $\mathbf{B}$ back to the system $\mathbf{S}$.

The population dynamics $P(t)$ change drastically depending on the ratio $\alpha :=  \omega_c/\Delta
$ (see Fig.~\ref{fig:Dataset}(c)). This parameter distinguishes between coherent and incoherent dynamics, and is therefore used, together with the ratio $\gamma/\Delta$, to categorize non-Markovianity in time evolution. For small values of $\alpha$, population oscillations occur over time; for larger values, however, a monotonic decay of the excited state is observed. $\alpha$ was logarithmically distributed within the range $\left[10^{-2}, 10^1\right]$, so the  $\log \alpha$ values were used as regression targets to facilitate prediction. 
The RFR, SVR and FFNN methods, which are discussed in Sec.~\ref{sec:Methods_ML}, were applied to this new regression task using these input and target sets. 
 The results using all three methods are shown in Fig.~\ref{fig:nonMark-Results}.
 
\begin{figure*}[ht]
    \centering
\includegraphics[width=\linewidth]{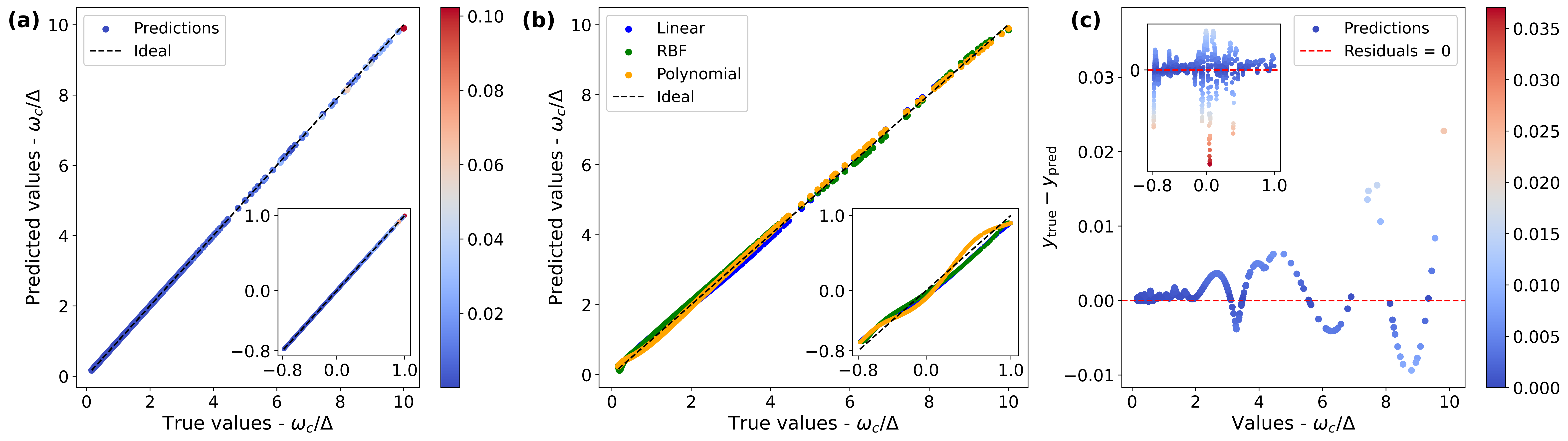}
    \caption{Regression results for the parameter $\alpha\equiv \omega_c/\Delta$. (a) RFR, comparing the actual values with the predicted values; the colorbar represents the squared absolute error. (b) Regression results using different types of kernels. (c) The error between actual and predicted data (residuals) is shown as a function of $\omega_c/\Delta$, and the colorbar represents the squared absolute error. In all cases, the insets show the results of the respective method on test data with a logarithmic scale.}
    \label{fig:nonMark-Results}
\end{figure*}
%%%%%%%%%%%%%%%%% SIGMA MEASURE  %%%%%%%%%%%%%%%%

Regressions were performed using three different machine learning methods, with targets including $\alpha%=\omega_c/\Delta
$, $\log \alpha$ and $\sigma_\alpha$. After the training process, the Mean Squared Error (MSE) and the $R^2$ score were used as error metrics on the test set. These values are summarized in Table~\ref{tab:nonMark_scores} for the different targets.
%%%%%%%%%%%%%%%%%%%%%
\begin{table}[h!]
\centering
\renewcommand{\arraystretch}{1.3}
\begin{tabular}{>{\centering\arraybackslash}p{0.2\linewidth}>{\centering\arraybackslash}p{0.25\linewidth}>{\centering\arraybackslash}p{0.25\linewidth}>{\centering\arraybackslash}p{0.25\linewidth}}
\hline
\multicolumn{1}{c}{\textbf{Labels}}   & \textbf{Method} & \textbf{MSE}   & \textbf{$R^2-$score}\\ \hline
                               & RFR               & $7.59211\times10^{-6}$ & 0.9999937 \\
                               & SVR - Linear      & 0.00568007 & 0.9952057 \\
$\alpha\equiv\frac{\omega_c}{\Delta}$       & SVR - RBF         & 0.00559125 & 0.9952807 \\
                               & RFR - Poly        & 0.00641122 & 0.9945886 \\
                               & FFNN              & $2.29932\times10^{-6}$ & 0.9999984 \\
\hline
                               & RFR               & $7.76782\times10^{-8}$ & 0.9999994 \\
                               & SVR - Linear      & 0.00657716 & 0.9520644 \\
$\log \alpha$                  & SVR - RBF         & 0.00577994 & 0.9578756 \\
                               & RFR - Poly        & 0.00623688 & 0.9545443 \\
                               & FFNN              & $3.76451\times10^{-8}$ & 0.9999997 \\
\hline
                               & RFR               & $1.82207\times10^{-9}$ & 0.9999998 \\
                               & SVR - Linear      & 0.00463015 & 0.6601655 \\
$\sigma_\alpha$                & SVR - RBF         & 0.00473560 & 0.6524264 \\
                               & RFR - Poly        & 0.00517200 & 0.6203962 \\
                               & FFNN              & $1.85983\times10^{-9}$ & 0.9999998 \\
\hline
\end{tabular}
\caption{Non-Markovianity regression metrics using different labels and the three ML methods discussed in this work (see Sec.~\ref{sec:Methods_ML}).}
\label{tab:nonMark_scores}
\end{table}
%%%%%%%%%%%%%%%%%%%

%%%%%%%%%%%%%%%%%%
\section{Discussion\label{sec:Discussion}}
We aim to fully characterize the properties of the environment through the time evolution of the components of the system  density matrix $\hat\rho_S$. Since the bath properties are set by the SD function contained in the correlation function $C_\beta(t)$, the determination of the parameters $s$, $\eta$, $\omega_c$ and $\gamma$ was sufficient for this task. A first dataset using $J_s(\omega)$ (Eq.~\eqref{Eq:Js(w)}) with
separate values of $s$ and $\eta$ was created to establish a scenario where the ohmicity and coupling strength regimes are easily identifiable from a physical perspective, as they exhibit sub(super)-damped oscillations in extreme cases. This task was successfully accomplished, achieving 100\% accuracy in the classification of $s$ on the test set and an MSE on the order of $10^{-10}$ for the $\eta$ regression, values significantly lower than those obtained for a uniformly distributed dataset. For the latter case, as shown in Table~\ref{tab:OhmicityScores_nonSep} and the confusion matrix in Fig.~\ref{fig:ConfusionMatrix}, the model also achieves 100\% accuracy in predicting the sub-Ohmic case (class 0),  characterized by a stronger damping of the coherence oscillations. For the remaining cases, however,  the model does not fully discriminate the transition between Ohmic and super-Ohmic regimes as these cases exhibit oscillations that extend beyond the given timescale. Nevertheless, these cases showed few false positives and false negatives.

To determine the value of $\eta$, the three most common regression methods in the literature were used~\cite{ML_probabilisticPersp_Murphy}. Each of these methods has its own advantages in terms of numerical implementation and underlying algorithmic logic. The RFR outperformed the other methods, achieving an overall error of the order of $10^{-9}$ and an $R^2$ score close to unity. In the case of the SVR, different types of kernels were used to determine which data mapping most efficiently captured the features of the quantum dynamics. Each kernel resulted in an MSE $\sim 10^{-2}$ which is significantly higher than the MSE obtained with RFR or FFNN.

Finally, a neural network was used due to its flexibility to be tuned  through various parameters, such as its architecture and training methods. In this case, a simple architecture, previously used in similar applications~\cite{Paternostro_1}, was adopted to avoid adding unnecessary complexity to the model. The accuracy of the neural network for regression or classification tasks also depends on the values of the learning rate, the batch size and the number of epochs. To standardize the training process, a preliminary characterization was performed by sweeping over different learning rates and batch sizes for a fixed number of epochs. 

The main parameters were selected based on their ability to achieve a convergent minimization of the cost function in fewer epochs. For the $\eta$ regression task, errors were within the same order of magnitude as those obtained with RFR, and a lower error was observed for non-separated data. These results demonstrate that the FFNN approach performs well in both cases, with a significantly low MSE, indicating a remarkable accuracy in predicting the values of $\eta$. Furthermore, the fact that the MSE is lower for non-separated data suggests that the model can efficiently handle more complex and less differentiated data distributions.

With this comprehensive analysis for the case of pure dephasing, it is evident that models have been established to recover the information related to the dissipative effect of the bath, encoded in the SD function, using only the evolution of the components $\hat\rho_{S}^{(ij)}$. This has been achieved without the need to explicitly calculate the dynamics for the test values or to have prior knowledge of the bath type. This is justified by the fact that a parametrization of the SD $J_s(\omega)$, as shown in Eq.~\eqref{Eq:Js(w)}, allows the approximation of different scenarios, such as Lorentz-Drude spectral densities by Ohmic approaches~\cite{SB_DynamicsOhmicBath} or underdamped spectral densities~\cite{UnderDamped_SpinBoson_HEOM}, which reproduce the dynamics previously studied~\cite{CaldeiraLegget_1, CaldeiraLeggett_2,JH_QDecoherence}. This lays the groundwork for the study of more complex spectral densities obtained experimentally or baths modeled as superpositions of multiple spectral densities with multiple peaks~\cite{MolecularSDreconstruction}.

%%%%%% MARKOVIAN REGIME  %%%%%%%%%%%%%%%
For the spin-boson system under consideration, Fig.~\ref{fig:Dataset}(c) shows that 
the values of $\omega_c/\Delta$ and $\gamma$ 
can be used to distinguish between Markovian and non-Markovian regimes, with the populations $P(t)$ decaying monotonically or oscillating over time. 
This suggests using these parameters to determine the non-Markovianity of the process by comparing them with known Markovian reference dynamics.
%%%%%%%%%%%
Our analysis covers the entire TLS-bath parameter space. By sweeping over the parameter $\Delta$, we covered both the weak coupling limit, $\gamma\ll \Delta$, and the strong coupling limit, $\gamma\sim \Delta$. We also covered all the regimes in terms of $\alpha$, analyzing the scaling limit, $\alpha\gg 1$, and  the crossover to the adiabatic limit, $\alpha \sim 1$. 

In addition, we have proposed a metric that uses the trace distance between operators to identify the flow of information from the environment to the system, Eq.~\eqref{eq:sigma_definition}. Although this may be somewhat related to the BLP metric~\cite{BLP_measure_Original}, which is sensitive to the initial state of the system, our label avoids the need to sweep over all possible initial conditions, a task that can be computationally intensive, as is the case when computing the BLP metric for arbitrary systems.
Our approach simplifies the regression task using three
different labels and lays the groundwork for determining
non-Markovianity in few-level systems. An extension to systems with more degrees of freedom~\cite{ML_nonMark_Dynamics} and more complex dynamics, where regimes are not as easily distinguishable
as in the spin-boson model, is a perspective of this work.

\section{Conclusions\label{sec:conclusion}}
We have demonstrated that, in a generic two-level system under the influence of a bosonic environment, spectral density function classification and regression can be performed with high accuracy. Starting from the baseline case of the spin-boson model under pure dephasing dynamics (i.e. no tunneling),  
we were able to contrast the model predictions with exact analytical solutions. For the spectral density (SD) Ohmicity classification, we achieved an accuracy above 98\%, which is consistent with previous studies~\cite{Paternostro_1}, but our approach used direct input from the time evolution and a larger dataset.
For the coupling strength %($\eta$)
from the regression analysis, we used the RFR, the SVR and the FFNN. We investigated the performance of all models, obtaining excellent regression results in all three cases, although RFR and FFNN outperformed SVR. With this classification and regression of the parameters $s$ and $\eta$ respectively, we were able to fully characterize the SD function and thus predict and identify  environmental effects from the system dynamics. 
%%%%%%%%%%
A crucial outcome of this work was generalizing the dynamics analysis to non-Markovian spin-boson systems using a numerically exact method to generate a new dataset. Using the numerical solutions generated for the general dissipative TLS, we successfully performed a regression analysis that quantified the degree of non-Markovianity using three different target labels. Once again, the RFR and FFNN models outperformed the SVR.

These results demonstrate the complete characterization of an unknown SD and the quantification of 
 non-Markovian dynamics using regression and classification approaches. We expect that these methods can be combined to characterize and control the environment under realistic conditions or from direct experimental data where no prior information on  possible non-Markovian effects is available. In particular, we expect these models to be useful in the study of quantum dissipation and noise spectroscopy of fermionic and spin environments.

\section{Data and Code Repository\label{sec:repo}}
\noindent
The database and Python codes used to develop this work are available upon reasonable request to the authors.
\section{Acknowledgements}
We  thank  the financial support of the Colombian Ministry of Science (MINCIENCIAS), Project No. BPIN 2022000100068, and the Norwegian Partnership Programme for Global Academic Cooperation (NORPART), Project NORPART-2021/10436 (QTECNOS).
%``Quantum Technology Education Consortium between Norway, South America and Cuba (QTECNOS)".  
J.H.R. and J.M.S. thank the Department of Physics at the University of Oslo, where part of this work was performed. 

\section{Appendix: Machine Learning methods\label{sec:Methods_ML}}

The FFNN algorithm for both classification and regression, together with the  RFR and SVR models for the implemented non-Markovian quantum noise spectroscopy characterization, are described below.

\subsection{Random Forest Regressor\label{sec:RFR}}
RFRs are part of ensemble methods that allow classification and regression tasks to be performed  based on the different features present in the data. These methods combine multiple weak learners (decision trees) and make decisions based on majority voting for prediction. These models avoid using a single tree with all the features, instead splitting them into different decision trees, thus reducing strong correlations between predictors and trees~\cite{DecisionTrees}. This ensemble approach addresses the overfitting problem that affects decision trees and also reduces the  variance in the fit.

For our dataset, the main features are the height
%intensity
of the oscillations and their damping, represented by the values of the density matrix $\rho_t^{(ij)}$ during all 200 time steps of the dynamics. Given that we have a dataset with a total of $n=10 000$ entries and a total of $p=200$ time steps (features), the bagging in the trees consists in randomly splitting the trees into $\sqrt{p}\approx 14$ features. For the regression case, majority averaging is implemented to fit
the values and make predictions on a test set. Figure~\ref{fig:RandomForest_sketch} outlines the implementation of the Random Forest.
%%%%%%%%%%%%%%%%%%%%%
\begin{figure}[h!]
    \centering
    \includegraphics[width=0.9\linewidth
    ]{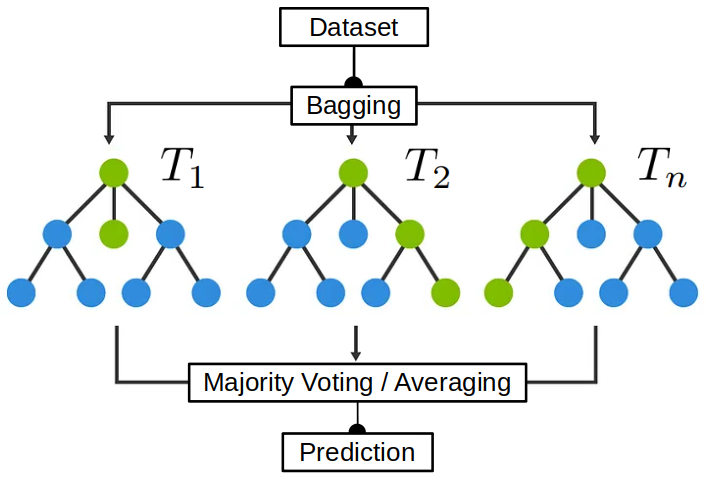}
    \caption{Sketch of the implementation of the Ensemble Random Forest Regressor with $n=100$ trees $T$ and majority averaging for the regression task.}
    \label{fig:RandomForest_sketch}
\end{figure}
%%%%%%%%%%%%%%%%%%

The architecture for the implementation of the RFR consisted of 100 \textit{number of estimators}, a \textit{minimum samples split} of 2 in order to prevent the data from splitting rapidly along each branch, and an undefined \textit{max depth}, 
so that each decision tree has as many branches as necessary until pure leaf nodes are obtained, or until all leaves contain fewer than \textit{min-samples-split} samples.

\subsection{Support Vector Regressor\label{sec:SVR}}

Support Vector Machine %(SVM) 
analysis is a popular machine learning tool for classification and regression tasks, developed by Vapnik \emph{et al.}~\cite{SVM_Vapnik}. This algorithm uses support vectors to identify and construct a hyperplane that best fits a given set of data points \cite{Vapnik_2018}. For regression problems, SVR is used  to find the optimal function $f(x)$ that separates the datasets and minimizes the distance between the targets and the predictions within a precision threshold $\epsilon$. Unlike traditional regression methods, which focus on minimizing the mean squared error, SVR ensures that the predictions remain within a predefined error margin $\epsilon$. This is achieved by optimizing a loss function that penalizes deviations greater than $\epsilon$, allowing greater control over the accuracy of the model. For datasets with more complex features, SVR can work efficiently  in higher dimensional spaces. This is because SVR can take  as a parameter a kernel that maps the feature space to a hyperdimensional space where finding a separating hyperplane is often asier~\cite{SVM_Vapnik}. Common kernels include linear, polynomial, and radial basis function (RBF). Each of these kernels transforms the data differently, allowing SVR to adapt to a wide variety of regression problems. In this work, an SVR is implemented  using the 200 features of dissipative quantum dynamics, and kernels such as linear, polynomial, and RBF are used for regression tasks. The choice of the appropriate kernel depends on the nature of the data and the specific problem to be solved.

\subsection{Feed Forward Neural Network\label{sec:FFNN}}

As a final method, an FFNN model was implemented, which takes the time evolution of the quantum dynamics as an input parameter and returns the regression value or classification, depending on the task, see for example Ref.~\cite{Goodfellow2016} for an in depth discussion of deep learning methods.

The network architecture consists of 200 units in the input layer, followed by hidden layers with sizes $[128, 64, 32, \text{out}]$, where $\text{out}=1$ for regression and $\text{out}=3$ for Ohmicity classification. The architecture is shown in Figure \ref{fig:FFNN_sketch}. The network was implemented in PyTorch \cite{Pytorch}  to perform supervised learning for regression and classification tasks. The former is trained by minimizing the cost function
%%%%%%%%%%%%
\begin{equation}
    C(y, \tilde{y}) = \sum_{k}^{N/\text{Bs}} \frac{1}{|M_k|} \sum_{i \in M_k} (y_i - \tilde{y}_i)^2  \label{eq:MSE_loss},
\end{equation}
%%%%%%%%%%%%%
where $y_i$ ($\tilde{y}_i$) is the target (predicted) value. The latter is achieved by minimizing the cross-entropy loss function
%%%%%%%%%%%%%%%%%%
\begin{equation}
    C(y, \tilde{y}) = - \sum_k^{N/\text{Bs}} \frac{1}{|M_k|}\sum_{i \in M_k} y_i \log \tilde{y}_i    ,  \label{eq:CE_loss}
\end{equation}
%%%%%%%%%%%%%%%%%%
where $M_k$ is the $k$-mini batch with size $\mathrm{Bs}=|M_k|$. The training process was performed using training and validation sets of size $\frac{3}{5}N$ and $\frac{1}{5}N$, respectively, where $N$ is the total number of data points. Training used 3000 epochs, batch sizes of 32 and 64, and learning rates between $10^{-4}$ and $10^{-6}$ to ensure proper convergence of the parameters.
%%%%%%%%%%%%%%%%%%%
\begin{figure}[h]
    \centering
    \includegraphics[width=0.9\linewidth
    ]{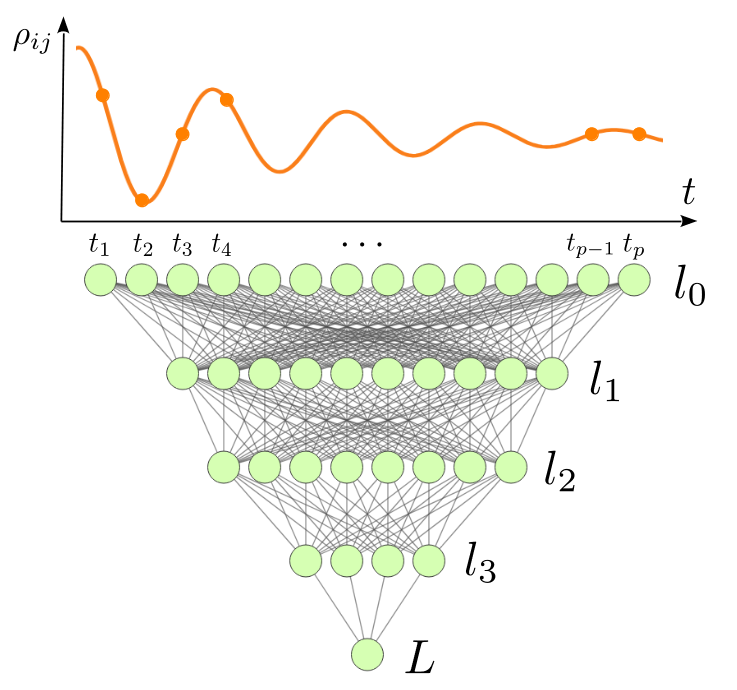}
    \caption{Architecture of the neural network used for regression and classification. The input layer $l_0$ consists of $p=200$ neurons, corresponding to the components of $\rho_{ij}$. The subsequent hidden layers have dimensions $[l_1, l_2, l_3] = [128, 64, 32]$, each with the ReLU($\cdot$) activation function. The output layer contains $L=3$ ($L=1$) neurons for the classification (regression) problem.}
    \label{fig:FFNN_sketch}
\end{figure}
%%%%%%%%%%%%%%%%%%%

For the classification task, $\text{out}=3$ since we have three classes and the output vector is passed through a \textit{softmax} function:
%%%%%%%%%%%%%%%%%%%
\begin{equation}
\sigma(x) = \frac{e^{x}}{\sum_i e^{x_i}}, \label{eq:softmax_func}
\end{equation}
%%%%%%%%%%%%%%%%%
where the exponential function is applied element-wise to the vector $x$. This function transforms the output vector into a probabilistic vector, such that the $i$-th component represents the probability of belonging to class $i$. On the other hand, for the regression task, $\text{out}=1$ and the activation function used is the ReLU function, defined as
%%%%%%%%%%%%%%%%
\begin{equation}
\text{ReLU}(x) = \max\lbrace x, 0\rbrace.  \label{eq:ReLU_activ}
\end{equation}
%%%%%%%%%%%%%%%%
This function ensures a positive output, which is suitable for predicting the values of $\eta$ and $\omega_c/\Delta$, since these parameters are always positive in the context of the problem. The network is trained to determine the optimal weights and biases $\theta=\lbrace W_i, b_i\rbrace$ that minimize the cost function given in Eq.~\eqref{eq:CE_loss}.  For this task, the Adam optimizer~\cite{AutomaticDiff_Pytorch} was used, which calculates the $n$-th gradient of the cost function $g^{(n)} = \nabla_\theta C\left(\theta^{(n)}\right)$ and the moments
%%%%%%%%%%%%%%%%
\begin{align}
& s \mapsfrom \rho_1 s + (1-\rho_1) \; g^{(n)} \nonumber \\
& r \mapsfrom \rho_2 r + (1-\rho_2) \; g^{(n)} \odot g^{(n)} \nonumber \\
& t \mapsfrom t+1 \nonumber \\
& \hat{s} = \frac{s}{1-\rho_1^t} , \qquad \hat{r} = \frac{r}{1-\rho_2^t}, \nonumber
\end{align}
%%%%%%%%%%%%%%
in order to follow the updating rule for the neural network parameters:
%%%%%%%%%%%%%%
\begin{equation}
\theta^{(n+1)} = \theta^{(n)} - \lambda \frac{\hat{s}}{\sqrt{\hat{r}}+\epsilon}, \label{eq:GD_ADAM}
\end{equation}
%%%%%%%%%%%%%%%%
where $\lambda$ is the learning rate, $\rho_1 = 0.9$, $\rho_2 = 0.999$, and $\epsilon = 10^{-8}$ for stabilization. After random initialization, the training process starts by following the recursive rules in Eq.~\eqref{eq:GD_ADAM} until the stopping criterion is met. This approach allows for efficient and robust optimization of the network parameters, ensuring fast and stable convergence during training.

\nocite{*}

\bibliography{apssamp}
\end{document}